# A One-Class Explainable AI Framework for Identification of Non-Stationary Concurrent False Data Injections in Nuclear Reactor Signals

Zachery Dahm[*], Vasileios Theos, Konstantinos Vasili, William Richards, Konstantinos Gkouliaras, and Stylianos Chatzidakis

**Affiliation:** School of Nuclear Engineering, Purdue University, West Lafayette, IN 47907

***Corresponding author:** zdahm@purdue.edu

**Abstract**

The transition of next generation advanced nuclear reactor systems from analog to fully digital instrumentation and control will necessitate robust mechanisms to safeguard against potential data integrity threats. One challenge is the real-time characterization of false data injections, which can mask sensor signals and potentially disrupt reactor control systems. While significant progress has been made in anomaly detection within reactor systems, potential false data injections have been shown to bypass conventional linear time-invariant state estimators and failure detectors based on statistical thresholds. The dynamic, nonlinear, multi-variate nature of sensor signals, combined with inherent noise and limited availability of real-world training data, makes the characterization of such threats and more importantly their differentiation from anticipated process anomalies particularly challenging. In this paper, we present an eXplainable AI (XAI) framework for identifying non-stationary concurrent replay attacks in nuclear reactor signals with minimal training data. The proposed framework leverages progress on recurrent neural networks and residual analysis coupled with a modified SHAP algorithm and rule-based correlations. The recurrent neural networks are trained only on normal operational data while for residual analysis we introduce an adaptive windowing technique to improve detection accuracy. We successfully benchmarked this framework on a real-world dataset from Purdue's nuclear reactor (PUR-1). We were able to detect false data injections with accuracy higher than 93% and less than 1% false positives, differentiate from expected process anomalies, and to identify the origin of the falsified signals.

**Keywords:** false data injection, anomaly detection, explainable AI, nuclear reactor

# Introduction

Advanced reactor designs, such as microreactors and small modular reactors, once implemented, will rely heavily on digital instrumentation and continuous live data monitoring to ensure safe and reliable operation (Coble et al., 2015; Lin et al., 2021; Ponciroli et al., 2016; Wood et al., 2017). Detecting and responding to cyber events in real-time is expected to play a critical role in safeguarding these systems, as undetected attacks could lead to inaccurate state estimations, costly operational downtimes, or even damage to critical components (Walker et al., 2022). Cyber threats, particularly false data injection (FDI) attacks, represent an area of concern for the nuclear industry, as highlighted by recent regulatory and safety frameworks (IAEA, 2011; Lawson-Jenkins, n.d.). The versatility and complexity of FDI attacks make them particularly insidious because they can mimic normal operating conditions or exploit correlations between multiple sensors, bypassing conventional threshold-based or correlation-driven anomaly detection systems (Mo and Sinopoli, 2009; Naha et al., 2023; Zhao and Smidts, 2020). They can involve simplistic approaches, such as injecting uniform false values, reproducing stationary steady-state signals, or even sophisticated strategies that generate false data closely resembling legitimate signals, e.g., replay attacks, to evade detection. The nuclear industry has already recorded instances of FDI-like attacks (Langner, 2011), underscoring the need for advanced detection and response mechanisms specifically tailored to address this threat.

FDI is a specific subclass of anomaly and therefore it is natural to implement tools from the field of anomaly detection (Blázquez-García et al., 2020; Choi et al., 2021; Schmidl et al., 2022; Zhao et al., 2021). Active monitoring (Willskyi, 1976; Zhai and Vamvoudakis, 2021) and watermarking (Li et al., 2018; Liu et al., 2020; Mo et al., 2015; Mo and Sinopoli, 2009; Satchidanandan and Kumar, 2017) are two methods for the specific detection of FDIs, which involve introducing small physical perturbations to the process data or adding a small amount of noise with a specific information pattern to the process data to verify that the current data has not been seen before. However, these methods involve modifying the process data in the reactor system which reduces control efficiency (Reda et al., 2022). It is worth noting the work of Mo and Sinopoli (2009) who analyzed replay attacks on control systems and mathematically proved that the asymptotic performance of a $\chi^2$ failure detector under a replay attack for a system based on linear time-invariant state estimation approaches the performance without attack making it possible for an adversary to evade detection. To address this, Mo and Sinopoli (2009) proposed the use of watermarking which increased the probability of detection at the expense of control performance. Zhao and Smidts (2020) further expanded this concept to include not only detection but classification of the anomaly into FDI or another anomaly by adding a second $\chi^2$ detector.

However, most of the approaches in literature rely on system models that are either linear time-invariant or with signal stationarity assumptions where the process is expected to have reached steady state which may not be representative of all modes of operation, e.g., load following.

While significant effort has been dedicated to anomaly detection using AI/ML (Muhlheim et al., 2023; Vilim and Ibarra, 2021) and optimization of watermarking signal variance (Ferrari and Teixeira, 2017; Liu et al., 2020; Mo et al., 2015; Mo and Sinopoli, 2009; Naha et al., 2023; Satchidanandan and Kumar, 2017) which, albeit increases in detection probability, has involved limited research on a path towards differentiation of FDIs from other anomalies with similar characteristics, e.g., equipment or sensor failure, number and origin of falsified signals, or interpretability of model detections which would be valuable for operator support and minimization of plant downtime (Zhao and Smidts, 2020). To meet stationarity assumptions and algorithm training requirements, FDIs often are generated using synthetic data by scaling or perturbing sensor values while at steady state (Ioannou et al., 2020; Reda et al., 2022) which is a valid approach and representative of some FDIs but is not necessarily the only possible vector for this attack. Collecting realistic data representing FDIs in nuclear reactors is exceedingly difficult due to the rarity of such events and the large number of combinations and ways they can be executed (Zhao and Smidts, 2020). In addition, the lack of real-world data precludes the wide implementation of these approaches as it is impossible to validate and test their performance, especially the false negative rates.

In this paper, we investigate the use of a passive AI/ML monitoring system combined with a modified SHAP analysis for identifying non-stationary concurrent replay attacks in nuclear reactor signals with minimal training data. We focus on nonstationary multivariate nuclear reactor time series with nonlinear and noisy characteristics. The proposed framework leverages recent progress on time series forecasting using recurrent neural networks and residual analysis which have been shown to perform better than other methods when it comes to anomaly detection in time-series (Schmidl et al., 2022). The recurrent neural networks are trained only on normal operational data while for residual analysis we introduce an adaptive windowing technique to improve detection accuracy. The original SHAP algorithm is modified to allow for interpretation of nonstationary signal contributions. Finally, a rule-based correlation combined with modified SHAP adds another layer of interpretation which helps identify the type of anomaly and number of falsified signals. We benchmark this framework on a real-world dataset from Purdue's nuclear reactor (PUR-1).

**Background**

The first new passive advanced nuclear systems are designed to be smaller—both in size and power output—than the commercial nuclear plants in operation today and will enable unattended operation in remote areas, e.g., microreactors and fission batteries. However, nuclear reactors today are based on technology developed for one-way power flow from a large power plant to passive customers at the receiving end. Advanced reactors, on the other hand, would likely be part of asymmetric networks and will require cyber-physical integration with communications infrastructure for remote sensing and control. Such infrastructure may include collection of information using sensors and devices from multiple sources that can include operations, transmission, distribution, electricity generation, customers, etc. (Holcomb et al., 2018;

Ramuhalli et al., 2018). A cyber-physical system is a system that integrates computation, networking, and physical processes designed to collect and analyze data from the physical world in real-time, and to use that data to control and monitor the behavior of physical processes (IAEA, 2011). Despite the obvious benefits of a cyber-physical system, the possibility of introducing new unintended vulnerabilities, previously non-existent due to the analog-based and proprietary nature of operation, cannot be ruled out (IAEA, 2011). This concern is reinforced by a steady rise in cyber-threats over the recent years. The nuclear industry has been a target of such cyber-threats for the past 30 years. This was ably demonstrated in attacks launched in 2014 against the Korea Hydro and Nuclear Power Co. (Baylon et al., 2015). Since the Stuxnet attack, many other incidents involved cyber-physical systems, including Duqu/Flame/Gauss (2011), Shamoon (2012), Havex (2013), Dragonfly (2014), Black Energy (2015) and Triton (2017). These events confirm that cyber-physical components are prime targets for both individual and nation-state actors (Zhang et al., 2020).

Cyber-attacks targeting power systems can generally be categorized into disruptions of availability, integrity, and confidentiality of information. Availability attacks result in inaccessible information due to communication interruptions, often achieved through denial-of-service (DoS) attacks. Integrity attacks involve the injection of incorrect information, commonly carried out using false data injections. Lastly, confidentiality breaches occur through unauthorized data access or usage, typically caused by brute force password cracking or stolen credentials. FDIs on the information layer manipulate the data that flows between sensors, controllers, and actuators, or alter the communication within a network such as a SCADA system. For instance, an attacker could inject fake temperature readings to cause the system to reduce coolant flow unnecessarily. However, because these manipulations target data rather than physical processes, the impacts may be delayed or subtle, making detection through conventional statistical approaches or other anomaly detection methods more difficult. In contrast, FDIs on the physical layer directly target the operation of physical components, such as pumps, valves, or reactor cores. These attacks involve manipulating data in a way that leads to physical changes in the system, such as altering actuator commands or modifying sensor data to drive the system into unsafe conditions.

A fundamental requirement of an FDI attack lies on intruder having access to network properties. Information "sniffing," typically required before gaining access to a network, will often trigger network security alarms (Zhang et al., 2019). However, once an intruder is connected in the system, tampering of transmitted information between communication parts is possible. Specifically, an attacker can alternate data in real-time by injecting prior captured data, stationary or non-stationary, without affecting communication consistency or triggering any network traffic alarms. Despite having low network traffic footprint, an FDI can introduce latency in data transmission which could be used as a potential detection signature (Theos et al., 2023). Currently, industrial control systems are safeguarded using perimeter defenses that rely on traditional Information Technology (IT) measures such as firewalls and data diodes or network traffic monitoring. While these efforts have the potential to improve the security of future nuclear communications, not all cyber-attacks are detectable by current intrusion prevention or detection systems that monitor network and host system data. Consequently, it is essential to investigate an additional layer of protection, known as physical process defense, to serve as a safeguard when traditional information security measures are compromised.

## Methodology

The proposed framework, shown in Figure 1, consists of four modules. The first module leverages recurrent neural networks and their sequence-learning capabilities to construct a predictive model of PUR-1 from eight reactor signals. The second module uses a residual analysis technique with adaptive short and medium window thresholding to identify potential anomalies in the data. The third module adds interpretability for error source identification by integrating a customized SHAP-based approach for time-series data. The last module adds another layer of interpretability by using rule-based correlations for differentiation between FDI and non-FDI anomalies.

### *Module 1: Predictive Modeling of Reactor Behavior*

The first step in the framework involves deploying an AI/ML predictive model trained exclusively on normal operational data to establish a baseline for expected reactor behavior. The goal is to capture the inherent dynamics of the reactor system by learning interdependencies between sensor signals and reactor states. By forecasting short-term future values of critical reactor parameters based on historical and real-time data, it provides a dynamic reference for normal operating conditions. Let $X(t) = [x_1(t), x_2(t), \ldots, x_n(t)]$ represent the set of $n$ sensor measurements at time $t$, where $x_i(t)$ is the reading from the $i$-th sensor. A predictive model $f$ is trained on historical data under normal operational conditions to learn the functional mapping:

$$\hat{X}(t + \tau) = f(X(t), X(t - 1), \ldots, X(t - k))$$

Where $\hat{X}(t + \tau)$ is the model's predicted state of the system at a future time $t+\tau$, based on the past $k$ time steps of observed data. This model uses a direct prediction scheme, so only the single value at time $t + \tau$ is predicted. The model is optimized to minimize the prediction error over normal operating conditions, such that:

$$\min_{f} \mathbb{E}\big[\|X(t + \tau) - \hat{X}(t + \tau)\|\big]$$

Where $\| \cdot \|$ represents an appropriate norm (e.g., mean squared error or absolute error), ensuring that the predicted values closely follow the true reactor behavior under normal conditions.

To optimize predictive accuracy and ensure adaptability across various reactor states—including startup, steady-state and transient power levels, and shutdown—multiple recurrent neural network (RNN) architectures were explored. RNNs were chosen due to their inherent ability to model dependent, multivariate time-series data, making them particularly well-suited for reactor monitoring, where sensor signals exhibit both strong interdependencies and temporal correlations. Previous research has demonstrated the effectiveness of time-series forecasting in nuclear reactor monitoring, particularly for predicting evolving reactor transients and accidents (Huang et al., 2021; Kaminski and Diab, 2024; Liu et al., 2015; Radaideh et al., 2020). Unlike traditional artificial neural networks (ANNs), which process inputs from different timesteps as independent features, RNNs introduce a hidden state mechanism that allows contextual information to persist across time steps, capturing sequential dependencies in the data.

Long Short-Term Memory (LSTM) networks were employed to address the limitations of standard RNNs in handling long-range dependencies (Hochreiter and Schmidhuber, 1997). LSTMs utilize two hidden states: a short-term memory state that evolves rapidly and a long-term cell state that retains information over extended sequences, mitigating issues of vanishing gradients. Additionally, Gated Recurrent Units (GRUs) were explored as a more computationally efficient alternative to LSTMs. By simplifying the memory architecture to a single hidden state, GRUs maintain competitive performance while reducing training time and computational overhead (Cho et al., 2014). This balance between expressiveness and efficiency makes GRUs particularly advantageous for real-time reactor monitoring applications.

*Module 2: Residual Analysis for Anomaly Detection*

The second module involves analyzing the residual errors of the predictive model from module #1. A key advantage of this approach is that the predictive model can be trained exclusively on normal operational data, which is significantly easier to obtain than labeled anomalous data. Additionally, unlike supervised learning methods, this technique enables the detection of previously unseen anomalies, making it particularly effective for identifying novel or adaptive attack strategies. The use of residual-based anomaly detection has been successfully demonstrated in (Ahmadi et al., 2023; Mahi-Al-rashid et al., 2022; Song and Ha, 2022), where residuals can be derived either from a reconstruction of the input data—commonly achieved using autoencoders (Li et al., 2022)—or from a time-series forecasting model that predicts future values of critical system parameters (Xu et al., 2022).

It is important to note that an FDI would remain undetectable by anomaly detection systems if the injected data closely mimics expected sensor values. However, such an injection is not considered successful unless it induces a harmful effect on system operations. The attack becomes effective once the falsified data deviates sufficiently from the true sensor values, influencing decision-making processes and potentially leading to unsafe or unintended system behavior. Let's assume that an adversary injects false data, resulting in a modified measurement vector:

$$\tilde{X}(t) = [\tilde{x}_1(t), \tilde{x}_2(t), \dots, \tilde{x}_n(t)]$$

where $\tilde{x}_i(t)$ represents the falsified value of the *i-th* sensor at time *t*. As a result, the deviation introduced by the FDI attack is given by:

$$\Delta X(t) = \tilde{X}(t) - X(t)$$

An FDI attack is undetectable by an anomaly detection system if:

$$\|\Delta X(t)\| \leq \epsilon, \forall t$$

for some small threshold $\epsilon$ that defines the sensitivity of the system. However, such an attack is not considered successful unless it induces a harmful effect on system operations. The attack becomes effective when:

$$\|\Delta X(t)\| > \epsilon_H, for\ some\ t$$

where $\epsilon_H$ is the minimum deviation required to cause a detrimental impact on system performance or safety. Thus, we arrive at a definition of what constitutes a successful FDI detection:

*Definition: An attack sequence is successfully detected when:*

$$\epsilon < \|\Delta X(t)\| \leq \epsilon_H, for\ some\ t$$

*indicating that the injected false data will be detected before causing any harmful consequences.*

In some contexts where a false data injection causes an extremely fast increase in $\Delta X(t)$, such as the false scrams presented in this work, the upper limit of $\epsilon_H$ may not apply as the deviation increases instantly to some very large value and the attack will need to be detected as soon as possible. Sections of the attack data where the discrepancy between the true values and injected false values is below $\epsilon$ are considered "undetectable" as they would not induce harm due to the false data being so similar to the true data. Module #2 continuously evaluates the residual errors —the discrepancies between predicted and observed reactor parameters. To identify the optimal threshold $\epsilon$, we employ a combination of adaptive thresholding and windowing techniques, enabling the identification of both subtle and sophisticated multi-sensor replay attacks.

*Module 3: Post-hoc Interpretability for Source Identification*

Detecting an anomaly alone does not confirm an FDI attack or identify the source, as deviations can arise from benign operational changes or sensor malfunctions. Additional interpretation is required to distinguish FDIs from other anomalies. In this module, we introduce a modified SHAP algorithm based on Shapley values (Lundberg and Lee, 2017) adapted for time-series data to assess whether the anomaly resulted from an FDI or another cause. If an FDI is identified, a correlation-based analysis is performed to further pinpoint the falsified sensors and characterize the nature of the attack. Unlike traditional interpretability methods such as Principal Component Analysis or Partial Dependence Plots, SHAP assigns individual feature contributions to each prediction, with feature contributions changing as inputs change. Prior studies have applied

Shapley values to determine global feature importance (Shin et al., 2023) and interpret outputs from general anomaly detection systems (Chaudhary et al., 2024; Fu et al., 2023; Park et al., 2022). The Shapley value for a given feature $j$ is:

$$\phi_j = \sum_{S \in \frac{\{1,\ldots,p\}}{\{j\}}} \frac{|S|!\,(p-|S|-1)!}{p!} (fun(S \cup \{j\}) - fun(S))$$

In this expression, $\phi_j$ represents the Shapley value assigned to feature $j$, which quantifies its contribution to the model's output. The summation runs over all or many possible subsets $S$ of features, excluding $j$. The total number of features in the dataset is denoted by $p$, while $|S|$ represents the number of features in subset $S$. The first term in the summation serves as a combinatorial weighting factor, ensuring that the contributions of $j$ are averaged across all possible subsets. The second term is a function which evaluates the model output based on only the subset $S$ and captures the marginal effect of adding feature $j$. The function $fun(S)$ is defined as:

$$fun(S) = \int f(x_1, \ldots, x_p) dP_{x \notin S} - E(f(X))$$

where $f(x_1,\ldots,x_p)$ is the model's prediction function. The integral term accounts for the expectation of the model's output when only the features in $S$ are considered, while the other features are marginalized over their probability distribution. The last term, $E(f(X))$, represents the expected value of the model's output over all feature inputs, which serves as a baseline reference. The sum of all Shapley values equals the difference between the actual output and $E(f(X))$, and each individual value represents the contribution of each feature to this difference. The SHAP algorithm computes the Shapley values for each feature more efficiently by creating a weighted linear model which predicts the model outputs for all possible sets of non-occluded features. Weights are added using the SHAP kernel, where $p$ is the total number of features and $|S|$ is the number of non-occluded features.

$$\pi_x(S) = \frac{p-1}{\binom{p}{|S|} |S|(p-|S|)}$$

Where $\pi_x(S)$ is the SHAP kernel weight, ensuring that each subset contributes fairly to the calculation. Using the SHAP kernel as a weight for the loss function of the linear regression model produces valid Shapley values as coefficients for each feature.

$$L(f, g, \pi_x) = \sum_{S \in S'} (f(S) - g(S))^2 \pi_x(S)$$

Where $L(f, g, \pi_x)$ represents the weighted loss function used to approximate the Shapley values. By minimizing the loss function the SHAP algorithm efficiently estimates Shapley values while maintaining computational feasibility, allowing it to scale to high-dimensional datasets.

## Module 4: Rule-based Anomaly Classification

A set of rules was developed to differentiate FDIs from non-FDIs by analyzing correlations between reactor signals and their respective rate signals. These rules use control rod direction, position, and neutron behavior to identify inconsistencies. For example, if a control rod moves but its recorded position does not align with expected motion, or if its movement direction does not match the measured position change, a correlation break is flagged. Similarly, neutron population trends are monitored, and deviations from expected change rates beyond a set threshold indicate anomalies. A neutron change rate rule further detects anomalies by identifying large variations in neutron count rates when no control rod movement occurs. If two consecutive valid data points show anomalies in any of these rules, the event cannot be from equipment malfunction or physics-based anomaly and is classified as an FDI. These correlation rules ensure that inconsistencies caused by data manipulation are flagged while minimizing false positives.

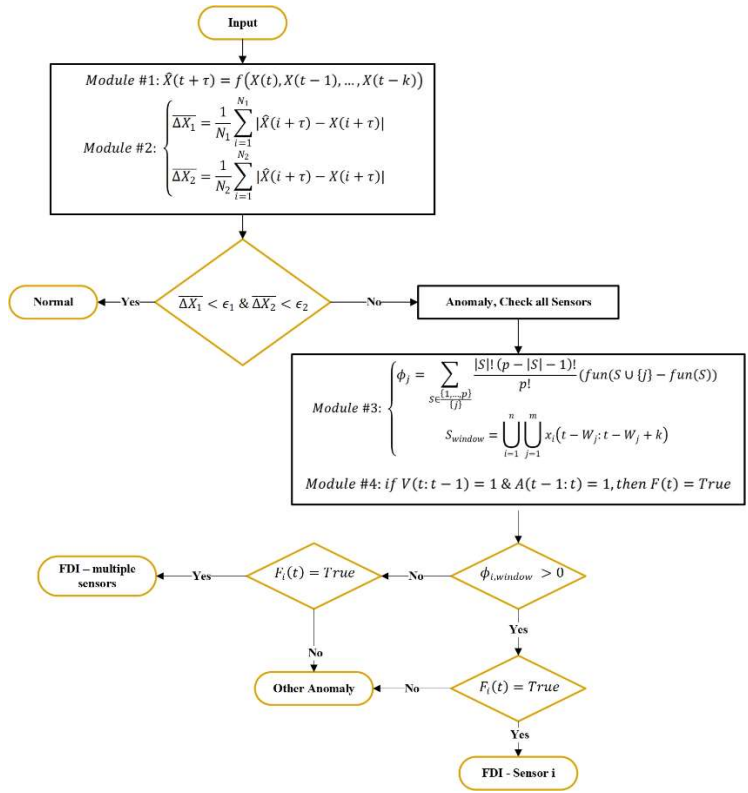

Figure 1: Framework Used for Identifying FDIs

# Use case

**PUR-1 system**

The use case is implemented in a real-world cyber-physical system by leveraging PUR-1, a research reactor located at Purdue University. PUR-1 is the only research reactor in the U.S. licensed with a fully digital instrumentation control system and a digital twin with two-way communications which allows collection of process, sensor, and IT data in real-time (Figure 2). PUR-1 is a pool-type research reactor licensed to operate at 10 kWth, and it is controlled through the movement of two control rods made of borated stainless steel and one regulating rod made of stainless steel.

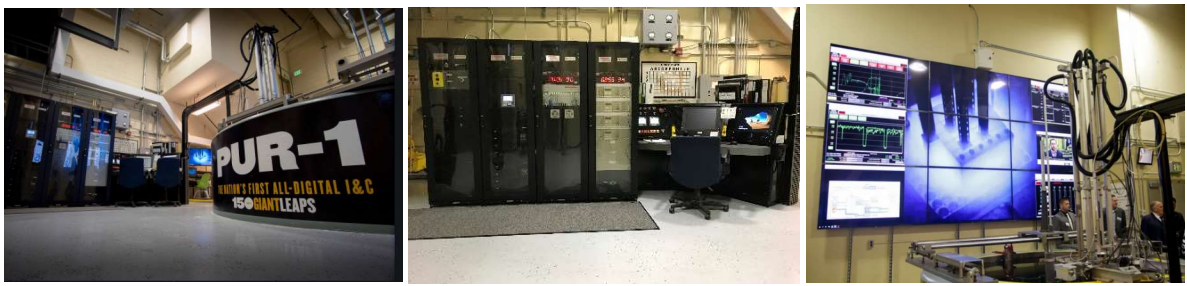

Figure 2: PUR-1 facility (left), fully digital control console (center), real-time data analytics

The cyber-physical system architecture includes five layers with various objects (e.g., equipment, components, or controllers), and relationships between objects (e.g., network connections with communication protocols and different signal flow requirements). Layers 0 to 3 are physically located in PUR-1 while Layer 4 is the outside network (also known as "business" or "remote") and is in a separate building. Layers 3 and 4 are connected via Ethernet/TCP-IP. IT security measures (e.g., firewalls or data diodes) are implemented between Layers 3 and 4. Physical OT signals and measurements for training, validation, and testing of AI/ML models are generated from Layers 0 to 3, while IT data is generated in Layers 3 and 4. All data collection, computations, and analysis (including model training, validation, and testing) takes place at Layer 4. Finally, an "attacker" PC is connected via an Ethernet switch to Layer 4 for introducing cyber events (e.g., DoS attack). An example of the Layering convention can be seen in Figure 3, along with possible entry points for cyber events.

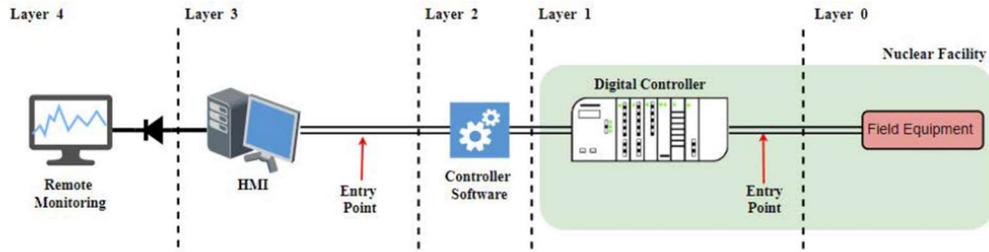

Figure 3: Architecture for injecting false data.

The following assumptions were considered when implementing the use case:

- It is assumed that an adversary has access down to Layer 1 (but not Layer 0).
- It is assumed that an adversary has no knowledge about the AI/ML system.
- It is assumed that the AI/ML models are located on a workstation in Layer 4 and that AI/ML model training and data processing takes place in a workstation located in Layer 4.
- It is assumed that an AI/ML model receives the same data and information as the reactor operators (Layer 3).

The use case involves injecting false data, aiming to deceive operators and automated systems into believing the reactor has shut down while it remains at high power. We selected five signals for falsification: neutron population, neutron population change rate, and the positions of the three control rods. These signals were chosen because they are typically the primary parameters monitored by operators on the control console and it is assumed that they will be the primary target for an adversary aiming to perform FDI. Figure 4 illustrates the system architecture developed for injecting and collecting false data and is divided into two primary sections: (i) Remote Location (Left Side) where the attacker PC connects to the monitoring workstation via a switch, allowing the injection of false data, and (ii) Control Room (Right Side) with the real-time reactor data network, including controllers, sensors, and HMIs, processes incoming signals for operational decision-making.

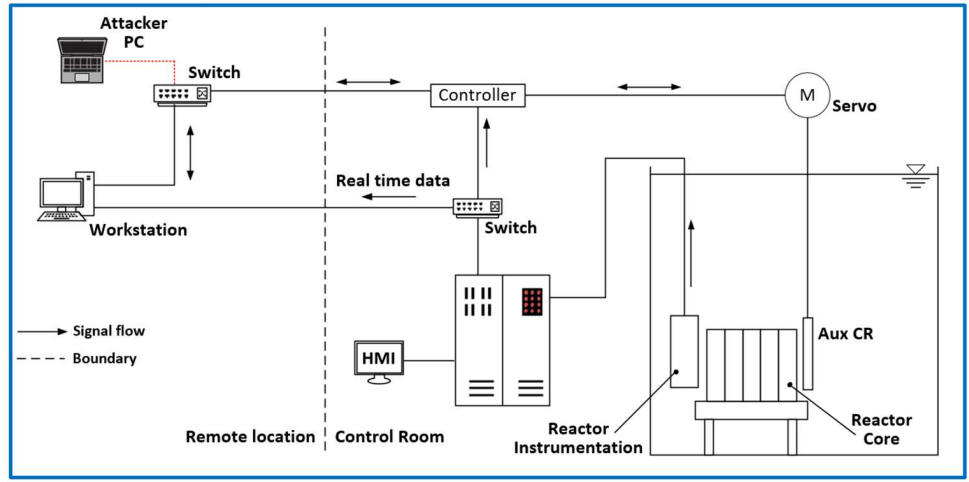

Figure 4: Architecture for injecting false data.

**Data collection**

All the data used to form the training and testing datasets is collected from PUR-1. 265,000 seconds of operation data were collected and processed between August 2022 and June 2023. Data was extracted using the R-time program and converted into a .csv file format with non-operating data filtered out. Then, 200,000 seconds of data were used to form the normal operation dataset, which was split into training, validation, and testing sets using a 60/20/20 split. The remaining 65,000 datapoints were used to form additional datasets with FDI and non-FDI abnormalities to further test and explore the performance of the proposed approach to unseen data. A total of six datasets were used in this experiment, derived from the original 265,000 points. The normal dataset #1 contains signals from 33 reactor operations (including startup, power increase/decrease, and shutdown) with average operation length of 5640 seconds, or approximately 1.5 hours each. Two more evaluation datasets (#2 and #3) were created to experimentally simulate challenging transient reactor behaviors that are not anomalies but may be detected as such by a conventional failure detection system. Three more evaluation datasets (#4, #5, and #6) were created to experimentally simulate FDIs of increasing complexity. These datasets are listed in Table 1 and shown in Figure 5.

The transient dataset #2 is a collection of 9 reactor transients, focusing on startup and power increases, where the absolute values of the neutron change rate are larger than 0.1 %/sec with a maximum of 5%/sec. The scram dataset #3 is a collection of 11 reactor scrams, with 60 seconds before and after the scram being included. In FDI dataset #4, the control rod positions, control rod active states, and neutron count change rate indicate that the reactor is operating at a consistent, high power, while the neutron counts are falsified and show the reactor scramming. The FDI #5 dataset is similar, but now both the neutron counts and the neutron count change rate are falsified and indicating a reactor scram. In the final FDI #6 dataset, it is assumed that an adversary has access to all monitored signals, i.e., neutron counts and change rate as well as control rod positions, except the control rod active states which are not actively monitored. The neutron counts, neutron count change rate, and three control rod positions are falsified and indicate that the reactor is shutting down while only the control rod active states are unaffected and show constant operation. Each false scram in the FDI dataset starts with 120 seconds of regular reactor operation data, then an injection of 70 seconds of neutron count data collected from a reactor shutdown, with a 10 second period of gradual change to the new neutron count value.

Table 1: Datasets derived from original full dataset of normal reactor operation

| Dataset | Use | Datapoints | Description |
| --- | --- | --- | --- |
| #1 Normal | Training, validation, testing | 200,000 | All reactor power cycles for one year, split into training, testing, and validation |
| #2 Transient (normal) | Testing | 19,800 | A collection of 9 transients where the absolute values of the neutron change rate is larger than 0.1 %/sec with a max of 5%/sec |
| #3 Scram (normal) | Testing | 1560 | A collection of 11 reactor scrams, including 60 seconds before and after the scram button is pressed |
| #4 FDI-A (abnormal) | Testing | 1560 | Injection of false data on one signal: neutron counts |
| #5 FDI-B (abnormal) | Testing | 1560 | Injection of false data on two signals: neutron counts and neutron change rate |
| #6 FDI-C (abnormal) | Testing | 1560 | Injection of false data on five signals: neutron counts, neutron change rate, and control rod positions |

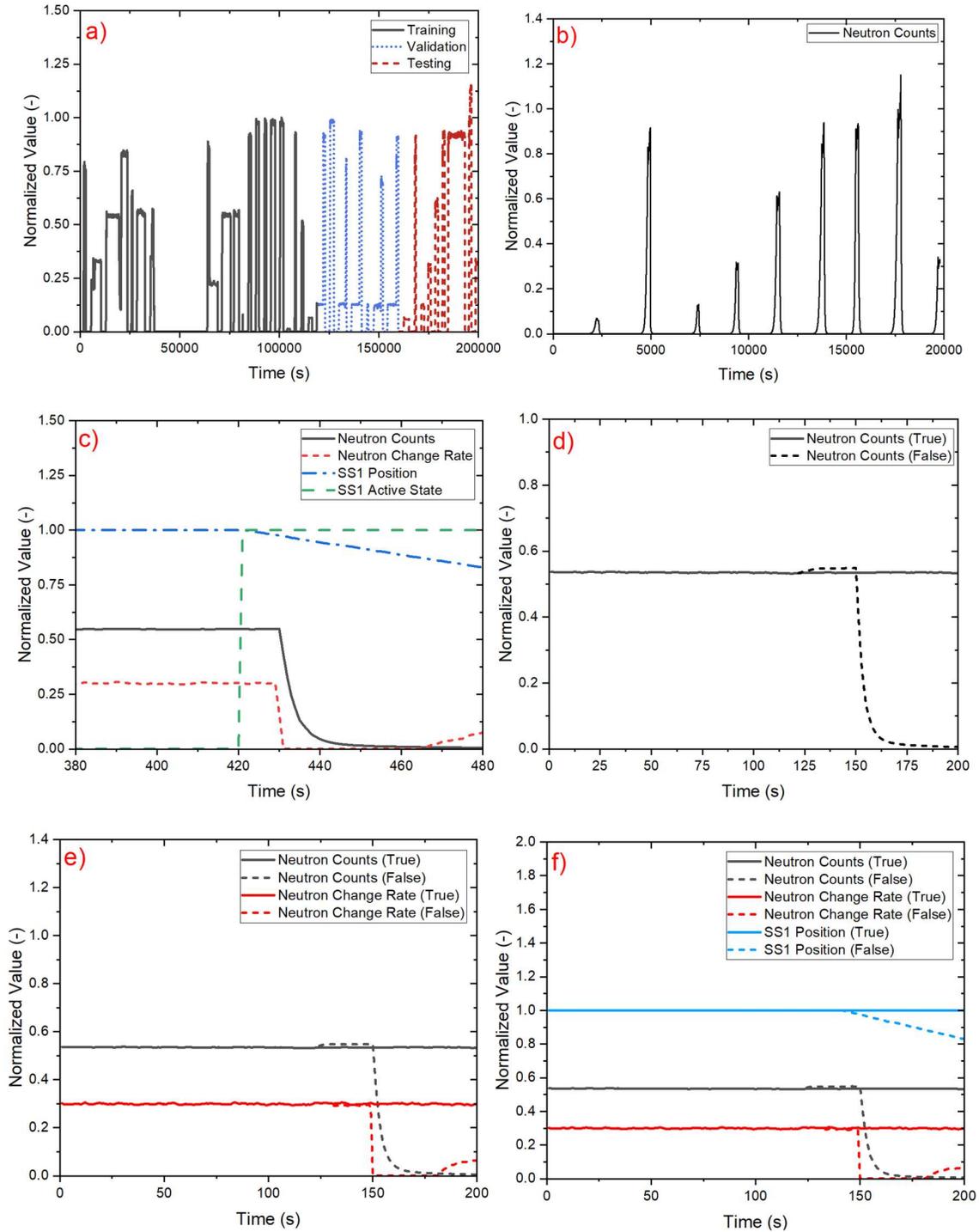

Figure 5: Full normal dataset containing training, validation, and testing sections (a); Neutron counts over time for the transient dataset (b). Example time series of Neutron Counts and Control Rod Positions for Scrams Dataset (c) and FDI-A dataset (d). Example False Scrams from FDI-B (e) and FDI-C (f) datasets.

**Data processing**

The constructed datasets underwent preprocessing to make them better suited for inputs to an AI/ML model (Figure 6). First, the data was normalized to a range between zero and one using a minmax scaler, which was fitted to the training dataset. All rows with any null values were dropped from all datasets. Then the data was split into time windows which are fed into each model, as shown in Figure 7. The windows were made to slide over the dataset one second at a time, regardless of length, so that a model output can be generated every second. Additionally, an index value from the original data collection was maintained for each datapoint and a check was performed to verify that the indices of all points and target in a time window were sequential to ensure that a time window that spanned over a discontinuity in the data was never used for training or evaluation.

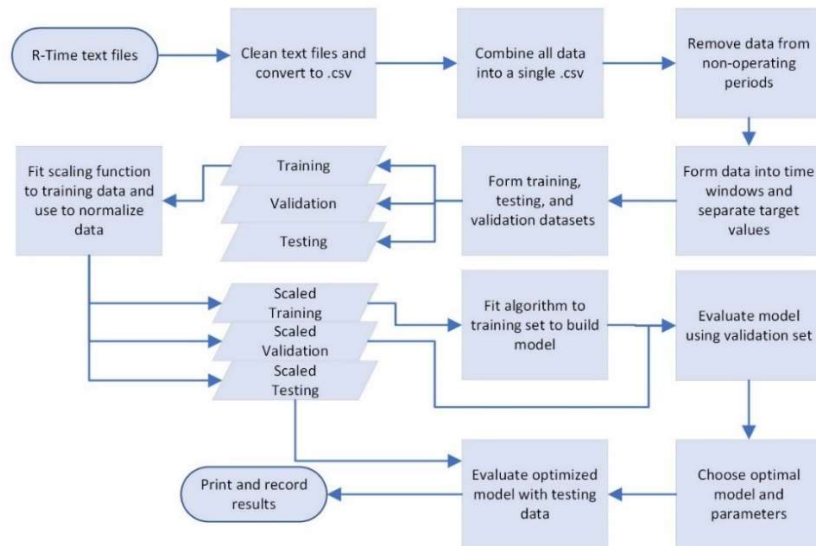

Figure 6: Flowchart showing preprocessing steps

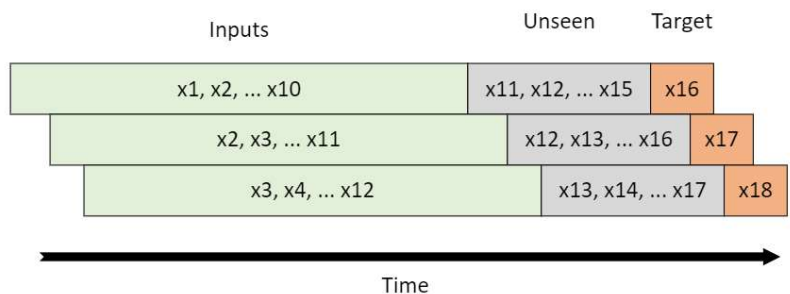

Figure 7: Example showing the sliding window approach and separation between inputs and target value

**Model construction**

Four different neural network architectures were trained, validated, and tested to find the best performing AI/ML model for module #1, i.e., predicting reactor neutron counts in the short term, and the model with the best performance was selected for more extensive evaluation. An artificial neural network (ANN), a recurrent neural network (RNN), a gated recurrent unit network (GRU), and a long short-term memory network (LSTM) were all trained and validated using the normal dataset (dataset #1).

*Tuning, training, implementation*

Table 2 shows the optimal set of hyperparameters found for each model, using 100 samples of the hyperparameter grid each. All the models performed the best with a single hidden layer, but otherwise the optimal hyperparameters differed significantly. During hyperparameter tuning, the models were trained using the normal training dataset and evaluated on the normal validation dataset. The criterion for selection of a parameter set was the lowest Mean Squared Error (MSE) on the validation dataset across the ten iterations of the model training:

$$MSE = \frac{1}{N} \sum_N (Predicted - Actual)^2$$

Table 2: Optimal Hyperparameters for Each Model Type

| Hyperparameter | ANN | RNN | GRU | LSTM |
|---|---|---|---|---|
| Learning Rate | 0.0001 | 0.001 | 0.0001 | 0.005 |
| Number of Stacked Hidden Layers | 1 | 1 | 1 | 1 |
| Neurons per Layer | 100 | 15 | 100 | 10 |
| Window Length | 10 | 30 | 30 | 10 |
| Number of Training Epochs | 5 | 50 | 50 | 20 |
| Batch Size | 16 | 8 | 4 | 8 |

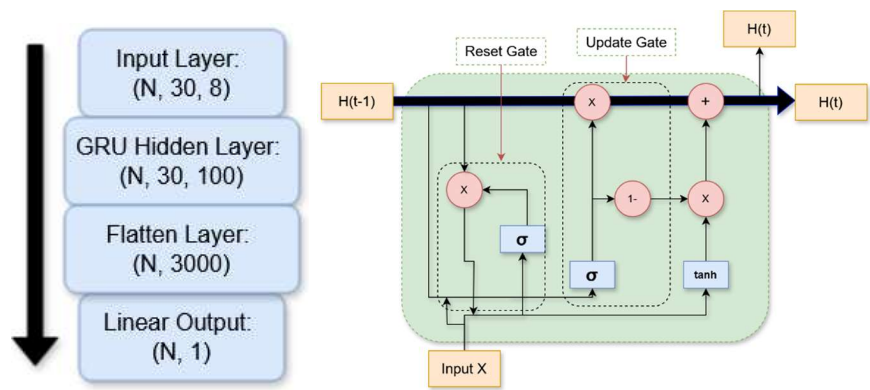

Figure 8: Diagram of GRU cell (Right) and GRU model (Left) showing layer output dimensions with batch size N

*Model Selection*

To gain an understanding of each model's performance on unseen data, all four models with their respective hyperparameters were applied to each of the five evaluation datasets (normal, transient, scram, FDI-A, and FDI-B) to see which model performs the best when predicting neutron counts five seconds into the future. The root mean squared error was used for model comparison:

$$RMSE = \sqrt{MSE}$$

Table 3 shows the RMSE for each model averaged across the entirety of each evaluation dataset. All models performed reasonably well, with average errors that were significantly higher as expected (above 0.15) on the FDI-A and FDI-B datasets than on the normal and transient behavior datasets (#1, #2, and #3) with errors ranging between 0.005-0.02. This highlights the ability of the models to differentiate between FDI and non-FDI events. GRU had the best performance on the normal and transient datasets, so it was chosen as the primary model to perform further analysis on. Figure 9 shows the training and validation loss for all the models as they trained for 500 epochs, which was done to ensure each model learned normally and validate the decision to use 50 epochs for GRU in the hyperparameter tuning process. GRU has the lowest validation loss by a visible margin, further reinforcing the decision to proceed with that algorithm. The training RMSE decreases slightly after 50 epochs but the validation RMSE slowly increases with larger numbers of epochs, so the use of 50 training epochs for the final model is reasonable.

A Python package, called Nuclear-data-TS-forecasting, accompanies this paper. It is publicly available at https://github.com/zachdahm/Nuclear-data-TS-forecasting and it comes with an example of running the proposed forecasting algorithms and the data used for the creation of the datasets. GRU training and predictions in Nuclear-data-TS-forecasting make use of the PyTorch, sklearn, and Pandas libraries for training and data processing. Simulations were run on an NVIDIA 3070 GPU with a training time of 46.6 minutes for 50 epochs for the GRU model.

Table 3: Averaged RMSE Across Full Datasets (Maximum error included in parenthesis)

|      | Testing  | Transient | Scram    | FDI #1   | FDI #2   |
|------|----------|-----------|----------|----------|----------|
| ANN  | 0.0138 (1.1119) | 0.0246 (0.2641) | 0.0179 (0.9352) | 0.1616 (0.7930) | 0.1487 (0.7456) |
| RNN  | 0.0113 (0.6944) | 0.0067 (0.1164) | 0.0137 (0.1381) | 0.1654 (0.7973) | 0.1533 (0.7494) |
| GRU  | 0.0085 (0.6606) | 0.0044 (0.1228) | 0.0184 (0.1691) | 0.1606 (0.7980) | 0.1463 (0.7499) |
| LSTM | 0.0141 (0.6751) | 0.0207 (0.4947) | 0.0147 (0.1558) | 0.1455 (0.7981) | 0.1336 (0.7497) |

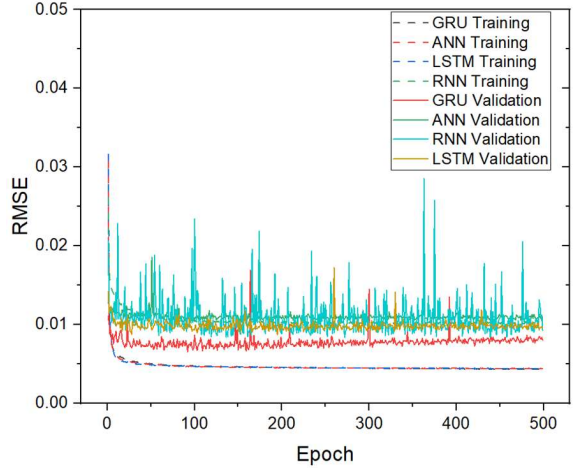

Figure 9: Training and Validation Loss Over 500 Training Epochs for All Models

**XAI Implementation**

*Modified SHAP Occlusion*

SHAP was originally designed to work with classifiers for tabular datasets. However, the computational cost of SHAP increases exponentially with the number of features, and in a direct application of SHAP to time series data, each feature at each timestep is treated separately, which can be too computationally expensive and can obfuscate overall trends through splitting contributions among too many timesteps. To circumvent this, a windowed approach called WindowSHAP can be used (Nayebi et al., 2023) where the individual feature time series is broken into windows, reducing the total number of features to test. The stationary window method introduced in that work was chosen due to its ease of implementation:

$$S_{window} = \bigcup_{i=1}^{n} \bigcup_{j=1}^{m} x_i(t - W_j : t - W_j + k)$$

In our use case, a further modification of WindowSHAP was needed as the model input is a small window of a much larger non-stationary time series with fast-evolving trends which are ultimately driven by operator actions. The typical SHAP method occludes data by replacing the "missing" values with the global average (Grechanuk et al., 2021) or randomly sampled points. This would not work well for this use case, because occluded values are meant to represent a baseline and expected behavior of the feature. In contrast, most periods of reactor operation are at a power level sufficiently far from the mean, and an instant change in power back to the global mean would not represent an expected baseline reactor behavior (Figure 10). To accurately describe the baseline condition of the reactor, which may change over time, a modified WindowSHAP was implemented with a moving baseline condition set for each of the 8 signals (features) as shown in Table 4. These conditions were chosen to represent the default state of the reactor: constant power with no control rod movements. Notably, this constant power represents the current operating power instead of a global baseline, so the neutron counts at the first second of the input time window are used to replace occluded sections of neutron counts in the input. The same is applied to control rod positions. Using this method, importance is attributed to the features whose evolution over the input time window causes the greatest change in error. Negative Shapley values indicate that the change in a certain feature is decreasing the error, indicating a correlation with the predicted signal, while positive Shapley values indicate that the change in an input feature is increasing the error, indicating a lack of correlation with the predicted signal. Shapley values near zero indicate that the signal is at the baseline or that the variation in a signal from the baseline did not influence the error.

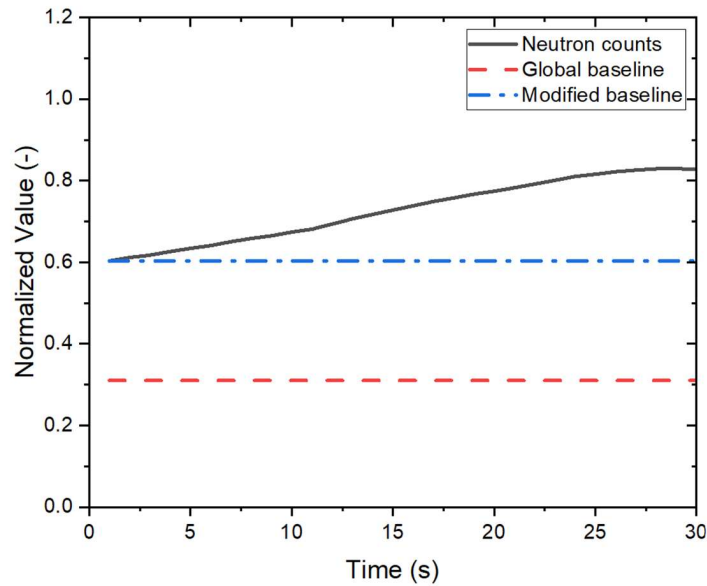

Figure 10: Comparison Between Power Values, Custom Moving Baseline, and Global Mean Baseline

Table 4: Baseline Condition for modified SHAP Occlusion

| Neutron Counts | Set to constant value from first second of input time window |
| Neutron Change Rate | Set to zero |
| CR Active States | Set to zero |
| CR Positions | Set to constant value from first second of input time window |

*Rule-based correlations*

A set of rules was implemented to analyze correlations between features. These rules are based on the fundamental relationships between control rod motion, neutron population changes, and their respective rate signals. A break in these correlations will indicate an FDI. We define the following notation: $D(t)$: Control rod movement direction (-1, 0, or 1) at time t; $P(t)$: Control rod position at time t; $A(t)$: Anomaly at time t; $V(t)$: Validity flag for the current data point; $N(t)$:

Neutron population at time t; CR(t): Measured neutron population change rate at time t; F(t): Final FDI determination.

*Control Rod Rule:* Control rod active states indicate whether a rod is moving and in which direction. If the measured control rod position does not align with its active state, this suggests a potential anomaly.

1. If the control rod position is greater than 3 cm and not NaN for the past two timesteps, mark the data point as valid:

$$if\ (P(t) > 3cm)\ \&\ (P(t:t-2) \neq NaN), then\ V(t) = 1$$

2. If the rod position increases by more than 0.7 cm but the movement direction is not positive, flag an anomaly:

$$if\ [(P(t) - P(t-1)) > 0.7\ \&\ D(t:t-2) \neq 1], then\ A(t) = 1$$

3. If the rod position decreases by more than 0.7 cm but the movement direction is not negative, flag an anomaly:

$$if\ [(P(t) - P(t-1)) < -0.7\ \&\ D(t:t-2) \neq -1], then\ A(t) = 1$$

4. If the position change is less than 0.7 cm but the movement direction is not zero, flag an anomaly:

$$if\ [abs(P(t) - P(t-1)) < 0.7\ \&\ D(t:t-2) \neq 0], then\ A(t) = 1$$

5. If two consecutive valid data points indicate anomalies, classify the event as an FDI:

$$if\ V(t:t-1) = 1\ \&\ A(t-1:t) = 1\ , then\ F(t) = True, else\ F(t) = False$$

*Neutron Change Rate and Neutron Counts Rule:* There is a well-defined relationship between neutron population and its rate of change.

1. If neutron population is above 1000, and no missing data exists for the last two timesteps, mark the data as valid:

$$if\ (N(t) > 1000\ \&\ CR(t:t-2) \neq NaN\ \&\ N(t:t-2) \neq NaN), then\ V(t) = 1$$

2. If the absolute difference between the expected change rate and the measured change rate exceeds 2, flag an anomaly. Expected change rate is calculated from the percent change in neutron counts in the last second, clipped to a minimum of -3:

$$if\ abs\left(\max\left(100 * \frac{N(t) - N(t-1)}{1 + N(t-1)}, -3\right)\right) - CR(t) > 2, then\ A(t) = 1$$

3. If two consecutive valid data points indicate anomalies, classify the event as an FDI:

$$if\ V(t:t-1) = 1\ \&\ A(t:t-1) = 1, then\ F(t) = True, else\ F(t) = False$$

*Neutron Change Rate and Rod Active State Rule:* Control rod movements are the primary driving factors for change in neutron population at PUR-1, so a large deviation in change rate without a control rod movement indicates an abnormality. A rule covering the opposite case of a control rod movement which results in a small deviation in change rate could not be effectively developed without very high false positive rates.

1. If previously shown validity conditions for neutron counts, neutron count change rate, and all three control rod positions are all true, mark the data as valid.
2. If the standard deviation of the neutron count change rate values over the last 10 seconds exceeds a threshold of 0.75, and all control rod movement directions for the last 10 seconds are zero, flag an anomaly
$$if\ all\ D((t-10):t == 0\ and\ \sigma_{CR(t-10:t)} > 0.75, then\ A(t) = 1$$
3. If two consecutive valid data points indicate anomalies, classify the event as an FDI
$$if\ V(t:t-1) = 1\ \&\ A(t:t-1) = 1, then\ F(t) = True, else\ F(t) = False$$

# Results

Averaged error values are typically used for demonstrating the performance of a forecasting model. However, this is insufficient in our use case because the performance over the full dataset is not yet known in real-time, and the goal for the model is real-time anomaly detection. Instead, we analyze the error for each time step in each evaluation dataset in real-time and label sections within each dataset as they are presented to the model. A model that produces too many false positives would not be useful as it contributes to alarm fatigue, and likewise a model that misses legitimate anomalies also cannot be relied upon to alert operators to abnormal situations. In our case, we use the mean absolute error:

$$MAE = \frac{1}{N}\sum_{N} |True - Predicted|$$

Figure 11 shows the optimized GRU model's performance when applied to the normal, transient, scram, and FDI-A, FDI-B, and FDI-C datasets, respectively, predicting neutron counts five seconds into the future. Across the normal, transient, and scram datasets, the model maintains a consistently low error, with some spikes in the normal dataset. Conversely, the model shows much larger error rates for the FDI datasets, meaning that disagreements between the neutron counts and control rod signals will lead to a high prediction error, which will in turn allow these disagreements to be identified as anomalies by the model.

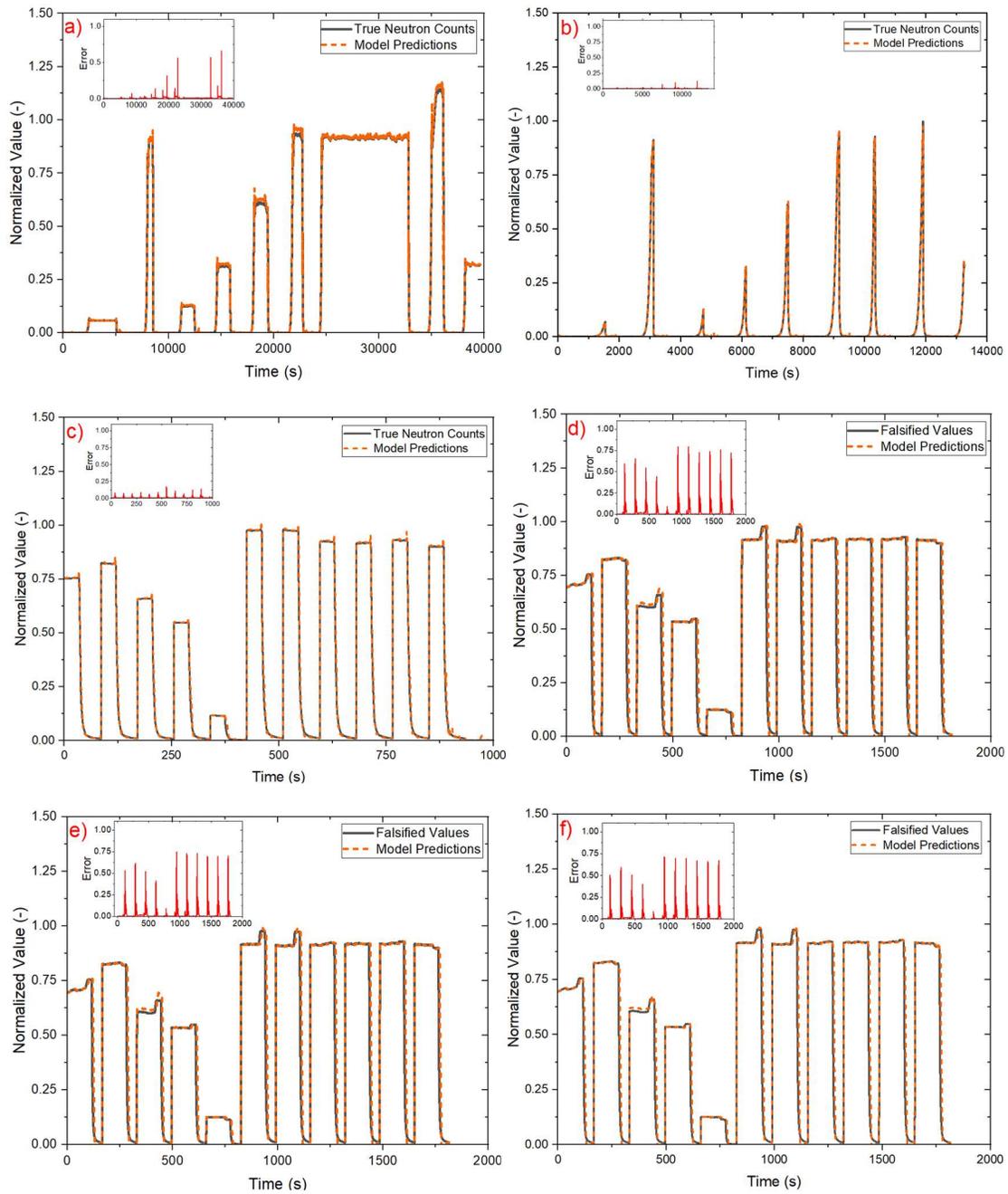

Figure 11: GRU Model Predictions and Error on Normal Dataset (a); GRU Model Predictions and Error on Transient Dataset (b); GRU Model Predictions and Error on Scrams Dataset (c) and FDI-A (d); GRU Model Predictions and Error on FDI -B (e) and FDI-C (f) Datasets.

First, the second-by-second anomaly detection rate in each dataset is determined, which shows the percentage of points in each dataset exceeding the error lower threshold. Viewing each prediction error individually and classifying FDIs based on a simple threshold is the most direct residual analysis method. Table 5 shows that this method has high accuracy (above 99% for 0.1 error threshold) on the normal, scram, and transient datasets. However, the performance when it comes to FDI datasets is between 66% to 76%. This is because not every point within the FDI dataset shows a large enough deviation from the expected behavior. Although performance appears low, this is not surprising given the difficulty of identifying these challenging FDIs. A different residual analysis method is needed to more comprehensively identify FDI attacks while remaining resilient to false positives.

Table 5: Per-second accuracy using error threshold

| Error Threshold | Normal | Scram | Transient | FDI-A | FDI-B | FDI-C |
|---|---|---|---|---|---|---|
| 0.01 | 0.8778 | 0.8597 | 0.9720 | 0.7614 | 0.7603 | 0.7548 |
| 0.05 | 0.9962 | 0.9620 | 0.9988 | 0.7680 | 0.7543 | 0.7581 |
| 0.1 | 0.9979 | 0.9928 | 0.9998 | 0.7300 | 0.7196 | 0.7157 |
| 0.15 | 0.9986 | 0.9982 | 1 | 0.6970 | 0.6876 | 0.6760 |
| 0.2 | 0.9988 | 1 | 1 | 0.6645 | 0.6595 | 0.6590 |

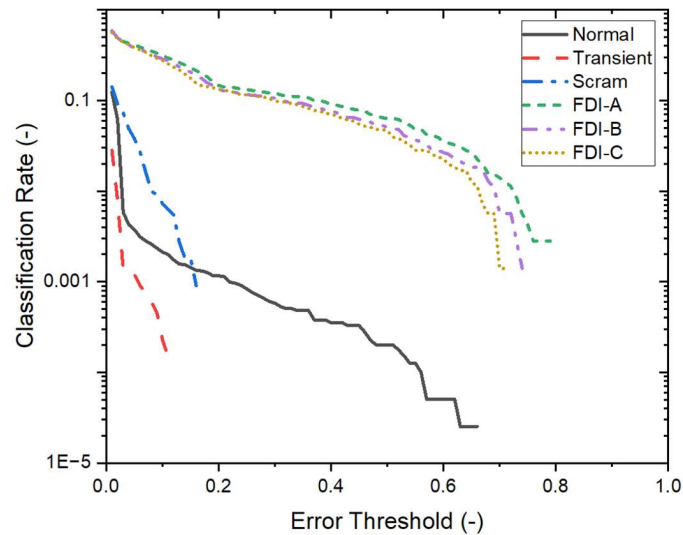

Figure 12: Relationship between the Error Threshold and the Proportion of Datapoints Labelled as Anomalies.

Figure 12 shows how the classification rate changes as a function of error threshold. To improve detection, we use an optimized combination of a short-term averaging window (error threshold 0.07 and 5 second window) and a medium-term thresholding window (error threshold 0.04 and 60 second window). Both windows calculate a rolling average of the residual over the last five or sixty seconds and compare this to the threshold. If the average exceeds either threshold, then the point is labelled as an anomaly. These two window lengths complement each other, as the short-term window is designed to detect large, sudden, and brief disturbances with very little delay. On the other hand, the long-term window is more sensitive at detecting FDI attacks over larger time windows and subtler attacks but will have some delay before an attack which is currently happening can sufficiently influence the lagging time window. Datapoints which exceed the short-term error window are passed to the XAI section for further qualitative analysis.

Table 6 shows that applying multiple rolling average windows improves accuracy on all datasets relative to the single per-second error threshold. The adaptive windowing method provides 93% accuracy in the FDI datasets and less than 1% false positives on all normal datasets. Once the power in the falsified data begins to drop, then the residual analysis method picks up the anomaly each time, with false positives usually being at the initiation of the FDI when false data is injected but before the false scram begins. This highlights the difficulty in using process data to detect a false data injection at the point of intrusion if there is not some large discontinuity in the data.

Table 6: Adaptive Window Residual Analysis Results on All Datasets

|  | Normal | Transient | Scrams | FDI-A | FDI-B | FDI-C |
|---|---|---|---|---|---|---|
| Accuracy | 0.9919 | 0.9993 | 0.9918 | 0.9348 | 0.9343 | 0.9348 |
| Precision | - | - | - | 1 | 1 | 1 |
| Recall | - | - | - | 0.8066 | 0.8045 | 0.8066 |
| F1 Score | - | - | - | 0.8929 | 0.8919 | 0.8929 |

Even though the proposed approach can differentiate between FDI-based anomalies and normally occurring transients or scrams it does not guarantee that an identified anomaly is an FDI. Further analysis and interpretation of the anomaly is needed. To achieve this, the SHAP outputs are analyzed using the modified WindowSHAP algorithm. In general, a large negative Shapley value for signal x indicates that the action in the predicted signal is correlated with the action in signal x. A large positive value indicates that the change in signal x is increasing the prediction error, showing a lack of agreement. A Shapley value near zero indicates that signal x is at the baseline behavior or that changes in this signal are not affecting the prediction error, which can indicate a lack of agreement. A pattern emerges from analyzing the SHAP outputs in the FDI datasets compared to the regular reactor signals, where in the FDI datasets the falsified signals have negative contributions to the error while the non-falsified signals have near zero contribution. This indicates that the false covariates in FDI datasets #4, #5, and #6 are linked to the false neutron counts. Signal values that come from the same timesteps will have negative contribution values, while unrelated signals will have positive or near zero contribution. This is shown in Figure 13 using the covariate of neutron count change rate. For the FDI datasets the points provided to the XAI section are the time windows which resulted in the highest residual, and for the scrams dataset, which had lower residuals, the end of the input time window was set five seconds after the scram, which roughly lined up with the positioning of the FDI time windows. The Shapley value of the final five-second time window within each input was averaged across all the shutdowns in each dataset each signal in Figure 14. Shutdowns in the FDI-A, FDI-B, and FDI-C datasets were false scrams, shutdowns in the scrams dataset were true reactor scrams, and shutdowns in the testing dataset were all non-scram shutdowns, conducted by gradually lowering the control rods instead of dropping them. The same shutdowns are included in the testing and transient datasets, so the transient data was excluded.

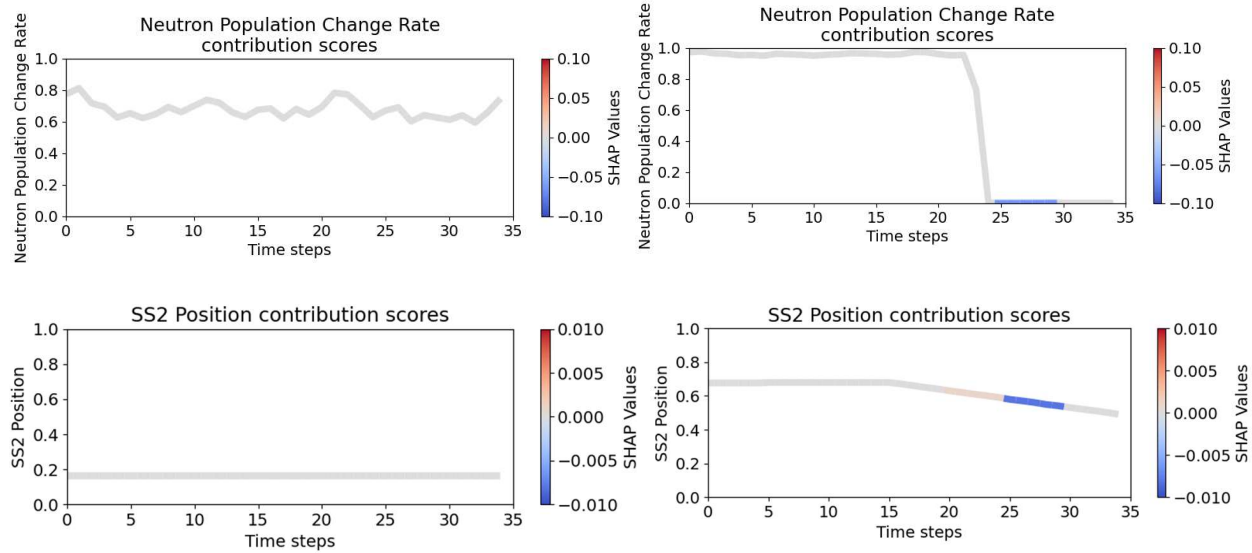

Figure 13: (Top Left) SHAP contributions for neutron count change rate in dataset where it is not falsified FDI-A (Top Right) SHAP contributions for neutron count change rate in dataset where it is falsified FDI-B (Bottom Left) SHAP contribution for SS2 position in dataset where it is not falsified FDI-A (Bottom Right) SHAP contribution for SS2 position in dataset where it is falsified FDI-A

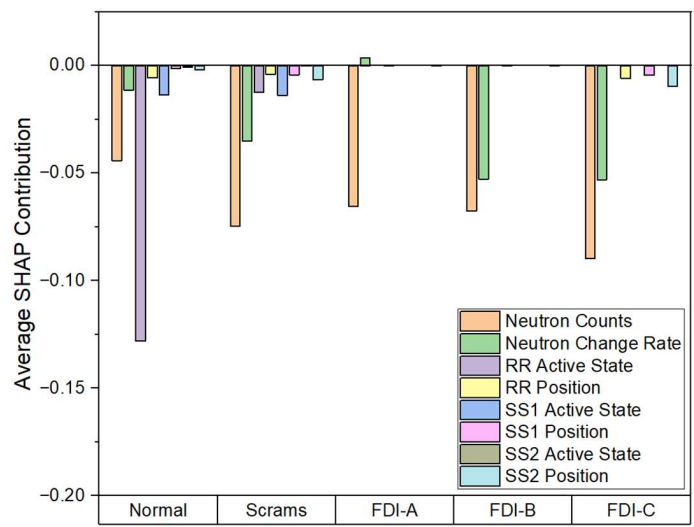

Figure 14: Averaged SHAP Feature Contributions for Each FDI Dataset

In addition to the SHAP explainability method, the rule-based approach was also implemented to identify which sensors are outputting faulty readings. The rules were developed from experience with PUR-1 data, and were made to target moments where the neutron change rate sensor significantly deviates from the hand-calculated change rate, when the control rod active state sensors significantly deviate from the actual change in control rod position, or when change rate has high variance without control rod movements. The correlations allow for differentiating the datasets but require knowledge of the dataset and possible exceptions to the correlation, and there are some relationships with too much variance to model with rules in PUR-1.

Table 7: Fraction of points identified as anomalies which break correlation rules in each dataset

|            | **Control Rods** | **Counts-Change Rate** | **Change Rate-Rods** |
|------------|------------------|------------------------|----------------------|
| Normal     | 0.005            | 0.008                  | 0                    |
| Transients | 0                | 0                      | 0                    |
| Scrams     | 0                | 0                      | 0                    |
| FDI-A      | 0                | 1.0                    | 0                    |
| FDI-B      | 0                | 0                      | 0.245                |
| FDI-C      | 0.940            | 0                      | 0.245                |

# Conclusions

In this paper, we evaluated a replay attack detection and identification framework based on AI/ML time-series forecasting combined with a customized SHAP algorithm while using one-class training data. The framework was benchmarked on real-world datasets from PUR-1 and we focused on multivariate nonstationary time-series to experimentally simulate replay attacks or more challenging normal reactor behaviors which ideally should still lead to low prediction error (scrams and transients). Four AI/ML algorithms were benchmarked against real-world normal datasets and datasets with artificial abnormalities and following hyperparameter tuning the results showed that all four models (ANN, RNN, LSTM and GRU) performed well with the GRU having the best performance. After using the GRU forecasting model to obtain residuals, errors were processed with an adaptive window thresholding technique and results showed 93% accuracy and less than 1% false positives on all normal datasets. The post-hoc customized SHAP interpretability method showed a pattern which helped to find the source of the falsified signals that caused a high residual error, and the rule-based approach provided a consistent and explainable method for identification correlation breaks between signals. The proposed approach can be applied to monitoring different sensors and would provide contextual information to operators about the origin of an issue based on which sensors are indicating unusual behavior. Future work would include a more comprehensive exploration of larger AI/ML architectures including transformers or hybrid/ensemble models, which could yield better results for forecasting and could make the framework more sensitive by decreasing residuals on normal data. Another piece of further work would be to generate additional types of anomalies or FDI use cases on different time scales and see how the model performs when fed new types of abnormal data, such as sensor drifting or scaling attacks. This could also prompt the creation of similar models which monitor other signals, which would help form a more comprehensive network of anomaly detection systems.

**Acknowledgements**

This research was performed using funding received from the DOE Office of Nuclear Energy's Nuclear Energy University Program under contract DE-NE0009268.